\documentclass[preprint,12pt]{elsarticle}

\usepackage{amssymb}
\usepackage{multirow}

\usepackage{amsmath,amssymb,amsfonts}%
\usepackage{amsthm}%
\usepackage{algorithm}%
\usepackage{algorithmicx}%
\usepackage{algpseudocode}%

\makeatletter
\let\elsarticle@keyword\keyword
\g@addto@macro\frontmatter{\let\keyword\elsarticle@keyword}
\makeatother
\usepackage{program}

\newtheorem{definition}{Definition}

\newtheorem{theorem}{Theorem}
\newtheorem{remark}{Remark}%

\begin{document}

\begin{frontmatter}



\title{NP-Completeness of Neighborhood Balanced Colorings}


\author{Saeed Asaeedi \corref{cor1}}

 \cortext[cor1]{Corresponding author}
\ead{asaeedi@kashanu.ac.ir}
\affiliation{organization={Department of Computer Science, Faculty of Mathematical Sciences, University of Kashan},
            city={Kashan},
            postcode={87317-53153}, 
            country={I. R. Iran}}


\begin{abstract}
A Neighborhood Balanced Coloring (NBC) of a graph is a red-blue coloring where each vertex has the same number of red and blue neighbors. This work proves that determining if a graph admits an NBC is NP-complete. We present a genetic algorithm to solve this problem, which we implemented and compared against exact and randomized algorithms.
\end{abstract}



\begin{keyword}
Neighborhood Balanced Coloring \sep NP-Completeness  \sep Vertex Coloring
\end{keyword}

\end{frontmatter}


\section{Introduction}
\label{intro}
Extensive research exists on edge and vertex labeling~\cite{gallian2012graph}. Cordial labeling was first introduced in~\cite{cahit1987cordial}, followed by balanced cordial labeling in~\cite{kaneria2016balanced}. A balanced cordial labeling of a graph $G=(V,E)$ is a cordial labeling $f:V \cup E \rightarrow \{0,1\}$ where the number of vertices and edges labeled 0 is equal to the number of vertices and edges labeled 1, respectively. Freyberg and Marr introduced \emph{Neighborhood Balanced Coloring} (NBC) in~\cite{freyberg2024neighborhood}. A graph admits an NBC if its vertices can be partitioned into two subsets such that each vertex has an equal number of open neighbors from each subset. While cordial labeling is a global property considering the entire graph, NBC is a local property focused on the immediate neighborhood of each vertex. This paper demonstrates that determining if a graph admits an NBC is NP-complete.

The remainder of this paper is organized as follows. Section~\ref{sec:2} proves the NP-completeness of the problem. Section~\ref{sec:3} presents an exact algorithm, a genetic algorithm, and a randomized algorithm for solving the problem, along with a discussion of their numerical results. Finally, Section~\ref{sec:4} concludes the paper by summarizing its achievements.

\section{NP-Completeness of the Problem}
\label{sec:2}

In this section, we consider the problem of determining if a simple, undirected graph $G = (V, E)$ admits an NBC. This section proves that this problem, referred to as the NBC problem, is NP-complete.

To prove the NP-completeness of the NBC problem, we use a reduction from the partition problem, which was shown to be NP-complete by Karp in 1972~\cite{karp1972reducibility}. The partition problem determines if a set $S$ of positive integers can be divided into two subsets such that both subsets have the same sum. We first define an $n$-pack, followed by a detailed presentation of the reduction steps.

\begin{figure}
\centering
  \includegraphics[width=0.55\linewidth]{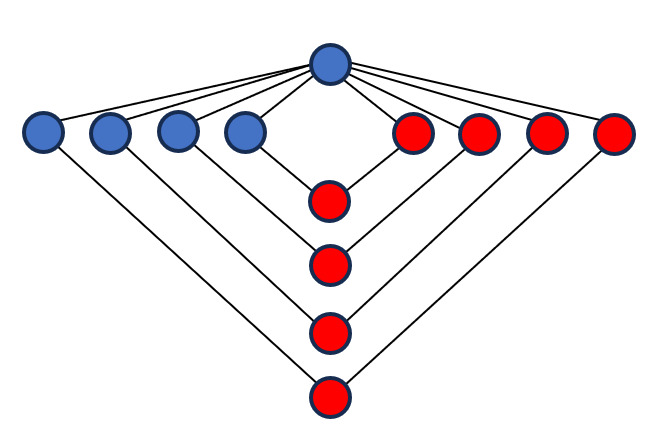}
\caption{In the $4$-pack graph, the top blue vertex represents the base vertex, the horizontal vertices represent support vertices, and the vertical red vertices represent numeric vertices.}
\label{fig:1}
\end{figure}

\begin{definition}
For $n>1$, an $n$-pack is a graph with $3n + 1$ vertices and $4n$ edges. The vertices consist of:
\begin{itemize}
\item
One base vertex: Connected to all 2n support vertices.
\item
$2n$ support vertices: Each connected to the base vertex and its corresponding numeric vertex.
\item
$n$ numeric vertices: Each connected to its corresponding two support vertices.
\end{itemize}
For n = 1, an n-pack is a graph with a single numeric vertex. 

(See Fig.~\ref{fig:1} for an illustration.)
\end{definition}

As depicted in Fig.~\ref{fig:1}, an $n$-pack admits an NBC. The following remark demonstrates that all numeric vertices in an $n$-pack must have the same color in any NBC.

\begin{remark}
\label{rmk:1}
In an $n$-pack, all numeric vertices must have the same color. This is determined by the base vertex. Since each support vertex has a degree of 2, the color of a numeric vertex is always opposite to the color of the base vertex. Therefore, if the base vertex is blue, all numeric vertices are red, and vice versa.
\end{remark}

It's important to note that the structure of an $n$-pack ensures that all numeric vertices are the same color. While changing the color of all (or some) support vertices to a different color creates another valid NBC, this doesn't affect the color of the numeric vertices. Therefore, the specific colors of the support vertices are irrelevant for determining the color of the numeric vertices.

We will now reduce the NBC problem to the partition problem. The following steps demonstrate how to construct a graph $G$ from a set $S=\{a_1, a_2, \dots, a_n\}$ of positive integers such that $G$ admits an NBC if and only if $S$ can be partitioned into two subsets with equal sums.

\begin{enumerate}
\item
Initialization: Start with the bipartite graph $K_{2,2}$ as the initial graph $G$.
\item
Adding $n$-packs: For each $a_i \in S$, add an $a_i$-pack to $G$.
\item
Connecting Numeric Vertices: Choose two non-adjacent vertices, $v_1$ and $v_2$, in the same part of the initial bipartite graph $K_{2,2}$. Connect both $v_1$ and $v_2$ to all numeric vertices of the added $n$-packs.
\end{enumerate}

\begin{figure}
\centering
  \includegraphics[width=0.85\linewidth]{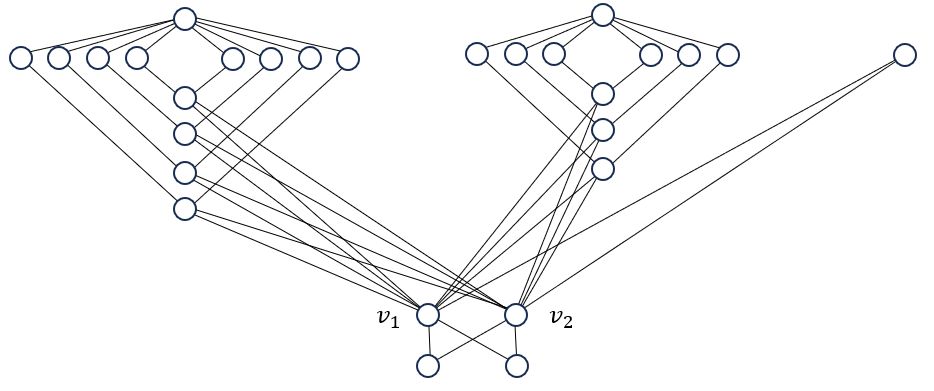}
\caption{The converted graph $G$, resulting from the reduction steps, includes the bipartite graph $K_{2,2}$, a $4$-pack, a $3$-pack, and a $1$-pack.}
\label{fig:2}       
\end{figure}

Fig.~\ref{fig:2} shows this conversion for $S=\{1, 4, 3\}$. The following theorem proves that $G$ admits an NBC if and only if $S$ can be partitioned into two subsets with equal sums.

\begin{theorem}
Given a set $S=\{a_1, a_2, \dots, a_n\}$ of positive integers, and let $G$ be the graph constructed using the presented reduction. Then, $G$ admits an NBC if and only if $S$ can be partitioned into subsets $S_1$ and $S_2$ such that $\sum_{a\in S_1} a =\sum_{a\in S_2} a$.
\end{theorem}

\begin{proof}
(if part): Assuming $G$ admits an NBC, vertex $v_1$ has an equal number of red and blue neighbors. Since $v_1$ is connected to all numeric vertices, half of these vertices must be blue and the other half red. According to Remark~\ref{rmk:1}, all numeric vertices in any $n$-pack are necessarily assigned the same color. Assume that the numeric vertices in the $r_1$-pack, $r_2$-pack, ..., $r_p$-pack are red, and the numeric vertices in the $b_1$-pack, $b_2$-pack, ..., $b_q$-pack are blue. This implies that $b_1+b_2+\dots + b_q = r_1+r_2+\dots + r_p$. Based on step 2 of the reduction, each $a_i$-pack corresponds to an item $a_i \in S$. Therefore, $S$ can be partitioned into two subsets $S_1=\{r_1, r_2, \dots, r_p\}$ and $S_2=\{b_1, b_2, \dots, b_q\}$, such that $\sum_{a\in S_1} a = \sum_{a\in S_2} a$.

(only if part): Assume $S$ can be partitioned into subsets $S_1$ and $S_2$ such that $\sum_{a\in S_1} a = \sum_{a\in S_2} a$. We will color the vertices of $G$ as follows:

\begin{enumerate}
\item
For each $n$-pack where $n\in S_1$, color the base vertex blue, the numeric vertices red, and half of the support vertices blue and the other half red. As previously explained, this coloring results in an NBC for this $n$-pack.
\item
For each $n$-pack where $n\in S_2$, color the base vertex red, the numeric vertices blue, and half of the support vertices blue and the other half red. As previously explained, this coloring results in an NBC for this $n$-pack.
\item
As desired, color vertex $v_1$ blue and vertex $v_2$ red. Their bottom vertices prevent them from being the same color. We also color them red and blue, respectively.
\end{enumerate}

This coloring scheme ensures that all $n$-packs admit an NBC. Since numeric vertices corresponding to items in $S_1$ are red and those corresponding to items in $S_2$ are blue, $v_1$ and $v_2$ are connected to an equal number of red and blue numeric vertices. Furthermore, each numeric vertex is connected to both $v_1$ and $v_2$ in addition to its connections to support vertices. As $v_1$ is blue, $v_2$ is red, and each $n$-pack is colored with an NBC, the entire graph $G$ admits an NBC.

\qed

\end{proof}

\begin{figure}
\centering
  \includegraphics[width=0.85\linewidth]{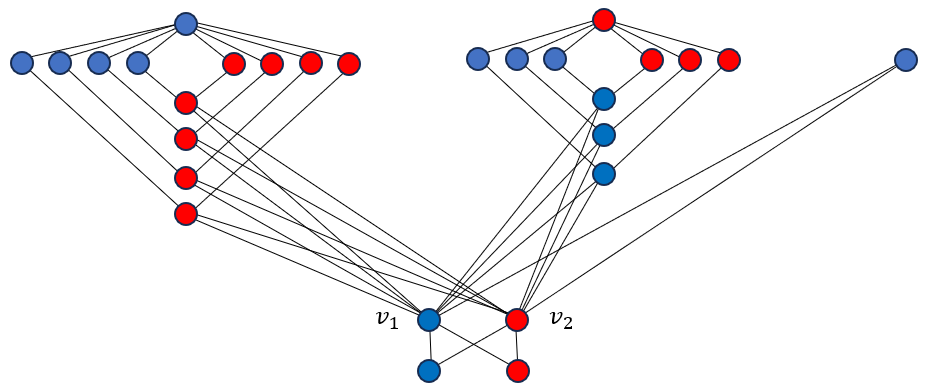}
\caption{The converted graph $G$, resulting from the reduction steps, includes the bipartite graph $K_{2,2}$, a $4$-pack, a $3$-pack, and a $1$-pack.}
\label{fig:3}       
\end{figure}

Fig.~\ref{fig:3} demonstrates how to color the corresponding graph $G$ for the set $S = \{1, 4, 3\}$. Notably, $G$ admits an NBC if and only if $S$ can be partitioned into $S_1 = \{4\}$ and $S_2 = \{1, 3\}$. As shown in Fig.~\ref{fig:3}, since $4=3+1$, numeric vertices of the $1$-pack and $3$-pack are colored blue, while those of the $4$-pack are colored red. The next section presents an exact algorithm and a genetic algorithm for finding an NBC (if one exists) for a given graph.

\section{Implementation and results}
\label{sec:3}

This section presents three algorithms designed to find an NBC. The brute-force algorithm (Algorithm~\ref{alg:1}) provides an exact solution with a time complexity of $O(n^2 2^n)$. In contrast, both the genetic algorithm (Algorithm~\ref{alg:2}) and the random coloring algorithm (Algorithm~\ref{alg:3}) have a time complexity of $O(n^2)$.

Since the input graphs for these algorithms may not admit an NBC, we aim to find a coloring that minimizes the penalty. The penalty for any given graph coloring is calculated as follows:

\begin{equation}
\label{eq:1}
penalty = \sum_{v\in V}{\mid r_v - b_v\mid}
\end{equation}

In Equation~\ref{eq:1}, $r_v$ and $b_v$ represent the number of red and blue neighbors of vertex $v$, respectively. Notably, the minimum penalty achievable across all possible colorings of a graph is 0 if and only if that graph admits an NBC.

\begin{figure}
\centering
  \includegraphics[width=0.8\linewidth]{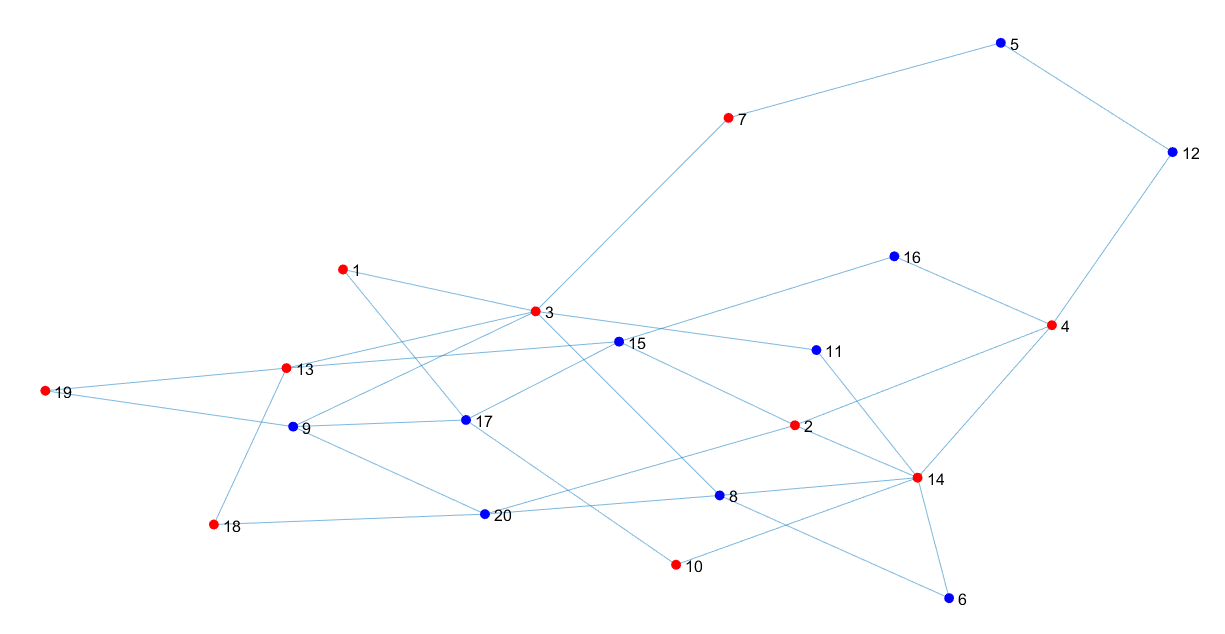}
\caption{The results of the genetic algorithm on a 20-vertex random graph.}
\label{fig:result}       
\end{figure}

Algorithm~\ref{alg:1} employs a brute-force approach, exhaustively exploring all possible colorings to find the one with the minimum penalty. Algorithm~\ref{alg:2} implements a single-crossover genetic algorithm, where each chromosome represents a coloring of the input graph. A '0' in the $i$-th bit of the chromosome corresponds to coloring the $i$-th vertex blue, while a '1' indicates red. Fig.~\ref{fig:result} illustrates the results of this algorithm on a randomly generated 20-vertex graph. Finally, Algorithm~\ref{alg:3} applies a random coloring to the vertices of the graph.

\begin{algorithm}
\caption{Exact Algorithm}\label{alg:1}
\begin{algorithmic}
\Require $G$ 
\Ensure $BestC$ as the vertex coloring with the minimum penalty
\State $n \gets$ number of vertices of $G$
\State $min \gets MAX\_INT$
\State $BestC \gets$ [ ]
\For{$i \gets 0$  to $2^n-1$ } \Comment{ Enumerate all possible colorings}
{
\State $C \gets$ The $n$-bit binary representation of $i$
\For{$j \gets 0$  to $n-1$ } 
{
\If{$C[j] == 0$ }
      \State Color vertex $V_j$ blue
\Else
\State Color vertex $V_j$ red
\EndIf 
} 
\EndFor

\State $Penalty \gets 0$

\For{$j \gets 0$ to $n-1$ }
\State $r_j \gets $ The number of red neighbors of vertex $V_j$
\State $b_j \gets $ The number of blue neighbors of vertex $V_j$
\State $Penalty \gets Penalty + \mid r_j - b_j \mid$ 
\EndFor

\If{$Penalty<min$}
      \State $min \gets Penalty$
      \State $BestC \gets C$
\EndIf 
} 
\EndFor

\end{algorithmic}
\end{algorithm}

\begin{algorithm}
\caption{Genetic Algorithm}\label{alg:2}
\begin{algorithmic}
\Require $G, ittCount, popSize, mutation\_rate$ 
\Ensure $BestC$ as the vertex coloring with the approximately minimum penalty
\State $n \gets$ number of vertices of $G$
\State $Pop \gets popSize$ random $n$-bit binary codes \Comment{Initialize population}

\For{$i \gets 1$  to $ittCount$ } 

\State $P_j \gets$ Calculate the penalty of the $j$-th member of $Pop$ using Equation~\ref{eq:1}.
\State  $S1 \gets$ Select the top $popSize/2$ members of $Pop$ with the lowest $P_j$ values.
\State  $S2 \gets$ Apply single-crossover operation to $popSize/2$ randomly selected pairs of chromosomes from $Pop$.
\State  $NewPop \gets S1 \cup S2$
\State  $NewPop \gets $ Mutate each chromosome in $NewPop$ with a probability of $mutation\_rate$.
\State  $Pop \gets NewPop$
\If{$min(P_j)==0$}
      \State break
\EndIf 
\EndFor

\State  $BestC \gets $ The chromosome in $Pop$ with the minimum penalty.

\end{algorithmic}
\end{algorithm}

\begin{algorithm}
\caption{Random Coloring Algorithm}\label{alg:3}
\begin{algorithmic}
\Require $G$ 
\Ensure $C$ as the random vertex coloring 
\State $n \gets$ number of vertices of $G$
\State $C \gets$ random $n$-bit binary code
\For{$j \gets 0$  to $n-1$ } 
{
\If{$C[j] == 0$ }
      \State Color vertex $V_j$ blue
\Else
\State Color vertex $V_j$ red
\EndIf 
} 
\EndFor

\end{algorithmic}
\end{algorithm}

\begin{figure}
\centering
  \includegraphics[width=0.75\linewidth]{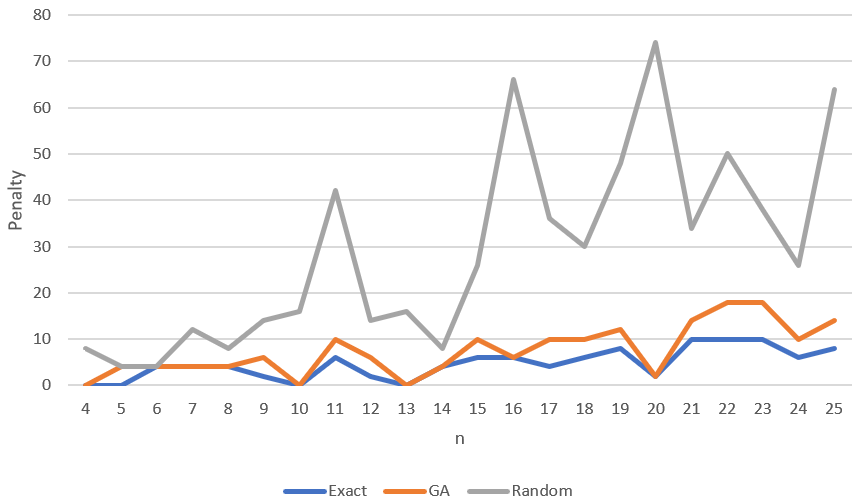}
\caption{Comparing the performance of exact algorithm (Exact), genetic algorithm (GA), and random coloring algorithm (Random) on graphs with varying sizes from $n=4$ to $n=25$.}
\label{fig:chart1}       
\end{figure}

\begin{figure}
\centering
  \includegraphics[width=0.75\linewidth]{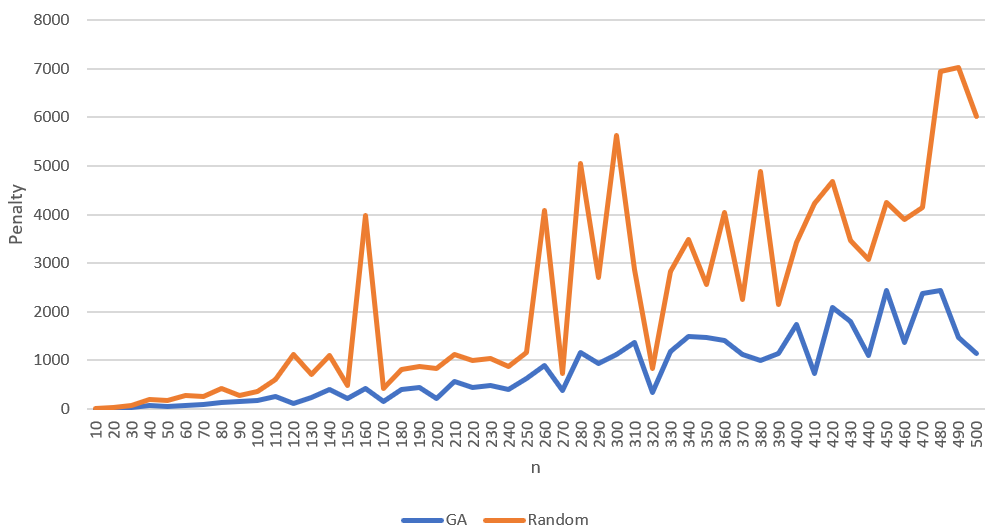}
\caption{Comparing the performance of genetic algorithm (GA) and random coloring algorithm (Random) on graphs with varying sizes from $n=10$ to $n=500$.}
\label{fig:chart2}       
\end{figure}

To evaluate the performance of these algorithms, we conducted comparisons on both small and large graphs. For small graphs (4 to 25 vertices), all three algorithms were evaluated. However, due to the time complexity of Algorithm~\ref{alg:1}, we compared only Algorithms~\ref{alg:2} and~\ref{alg:3} on larger graphs (10 to 500 vertices). Fig.~\ref{fig:chart1} presents the results for the three algorithms on small random graphs, while Fig.~\ref{fig:chart2} showcases the performance of Algorithms~\ref{alg:2} and~\ref{alg:3} on larger random graphs. 

\section{Conclusion}
\label{sec:4}

Here, the NP-completeness of determining whether a given graph admits an NBC is proved. To address this computational challenge, three algorithms are presented. First, a time-consuming but exact algorithm is developed and implemented on small-scale datasets. Second, a single-crossover genetic algorithm is designed and implemented on both small and large-scale datasets to address the computational challenges associated with the exact algorithm. Finally, a random coloring algorithm serves as a baseline for evaluating the efficiency of the genetic algorithm on both small and large-scale datasets.

%
%
%
%





\begin{thebibliography}{00}

\bibitem{gallian2012graph}Gallian, J. Graph labeling. {\em The Electronic Journal Of Combinatorics}. pp. DS6-Dec (2012)

\bibitem{cahit1987cordial}Cahit, I. Cordial graphs-a weaker version of graceful and harmonious graphs. {\em Ars Combinatoria}. \textbf{23} pp. 201-207 (1987)

\bibitem{kaneria2016balanced}Kaneria, V., Patadiya, K. \& Teraiya, J. Balanced cordial labeling and its application to produce new cordial families. {\em Rn}. \textbf{55} pp. 7 (2016)

\bibitem{freyberg2024neighborhood}Freyberg, B. \& Marr, A. Neighborhood Balanced Colorings of Graphs. {\em Graphs And Combinatorics}. \textbf{40}, 41 (2024)


\bibitem{karp1972reducibility}Karp, R. Reducibility Among Combinatorial Problems. {\em Proceedings Of A Symposium On The Complexity Of Computer Computations, Held March 20-22, 1972, At The IBM Thomas J. Watson Research Center, Yorktown Heights, New York, USA}. pp. 85-103 (1972), https://doi.org/10.1007/978-1-4684-2001-2



\end{thebibliography}


\end{document}